\documentstyle[12pt]{article}
\def\be{b}
\def\comment#1{}
 \newcommand{\sbf}[1]{\mbox{\scriptsize\bf{#1}}}

\title{\vspace{-1.5cm}
\Large \bf Universality Principle for Orbital Angular
Momentum and Spin in Gravity with Torsion
}
\author{ H. Kleinert        \\
        {\footnotesize  Institut f\"ur Theoretische Physik}\vspace{-2.5mm} \\
        {\footnotesize  Freie Universit\"at Berlin} \vspace{-2.5mm}\\
        {\footnotesize  Arnimallee 14 \ \ \ \ \ D 14195 Berlin, Germany}%
\vspace{-5mm}%
\thanks{kleinert@physik.fu-berlin.de ,
~http://www.physik.fu-berlin.de/\~{}kleinert \hfil
}
 }
\date{\footnotesize \today
        \vspace{-10mm}}

\begin{document}

\maketitle
\vspace{1cm}
\begin{abstract}
We argue that compatibility
with
elementary particle physics requires
gravitational theories with torsion
to be unable to distinguish
between orbital angular momentum and spin.
An important consequence
of this principle is
that spinless particles must move
along autoparallel trajectories, not along
geodesics.
\end{abstract}
~\\
{\bf 1.~}Universality principles provide us with
 important guidelines for constructing
candidates for
fundamental theories which have a chance of being true.
For example, an essential property of Maxwell's theory
is that electromagnetic interactions
depend only on the charge of a particle, not on
the various physical origins of charge.
The charge of an ion is composed of electron and nuclear charges, the
latter of proton charges, these in turn of quark charges,
which eventually me turn out to arise
from further charged substructures. The motion of a charged particle in an
electromagnetic field
does not depend on these details, which are
subject to change by
future discoveries.
An atom moves like
a neutral point particle,
in spite of the
completely different origins
of electron and proton charges,
the exact neutrality of an atom
being the very basis for
the electrostatic stability of large gravitational
bodies (and thus for the existence of theoretical physics).

The irrelevance of the physical origin of the ``charge"
of gravitational interactions, the  mass,
led Einstein to the discovery of a geometric theory
of these interactions. Just as electric charge, also the mass
of a
particle
has a multitude of origins, arising from
the masses of constituents and
 various field energies holding these together.
Gravitational interactions depend only on the
total mass, and  this property
makes all particles run along the same trajectories,
which can therefore be used to define a geometry of spacetime.
In Einstein's theory the independence of the physical origin of the
mass is ensured by the fact that the
 Einstein curvature tensor
is directly proportional to the
total energy momentum tensor of the theory.
Its
precise composition depends on the actual status of
elementary particle physics, but
the motion is invariant with respect to
this composition, and thus to future
discoveries on
the internal structure of the particles.

The universality of weak and color charges
was an important principle in the construction of
  unified theories of electromagnetic and weak, as well as of
strong interactions.
~\\
~\\
{\bf 2.~}Since a number of years, theoreticians have enjoyed the idea
that the geometry of
spacetime may not only be curved but carry also
torsion.
The line of arguments
leading to this idea
was
that Einstein's gravity
may be viewed as a gauge theory
of local translations.
These generalize
the
global translations under which all local
theories are invariant in Minkowski spacetime.
But the latter theories are also invariant under
the larger
Poincar\'e group, the group of translations and Lorentz transformations.
It therefore seemed natural    to
postulate the existence of a second
gauge field which ensures the invariance under
local Lorentz transformations \cite{hehl}.
As a result one obtains an interaction
between torsion and  spin
\cite{GFCM}
  in a four-dimensional spacetime with general coordinates $q^\mu$:
\begin{eqnarray}
{\cal A}=-\frac{ 1 }{2 }\int d^4q \sqrt{-g} K_{\mu \nu  \lambda } \Sigma ^{ \nu  \lambda,\,\mu},
\label{@inta}\end{eqnarray}
where
$
K_{\mu \nu  \lambda }$
is the contortion tensor, containing the torsion in the combination
$K_{\mu \nu  \lambda }=
S_{\mu \nu  \lambda }
-S_{ \nu  \lambda \mu}
+S_{ \lambda \mu \nu  \ }$. The tensor
$\Sigma ^{ \nu \lambda ,\,\mu}$ is
the local spin current density.

Consider now a particle at rest in a  Riemann-flat space
with euclidean coordinates
${\bf x}=(x^1,x^2,x^3)$ with $x^i=q^i$ for  $(i=1,2,3)$,
and a time $t=q^0$.
We assume the presence of a torsion whose only nonzero
components are
\begin{equation}
S_{ij0}=\frac{1}{2} \epsilon _{ijk}\be_k,~~~\partial _i\be_i=0,
\label{@SB}\end{equation}
The
divergenceless vector ${\bf b}$ will
at first be assumed to be constant, for simplicity of the argument.
Then (\ref{@inta})
specifies an interaction energy
\begin{equation}
H_{\rm int}=-\frac{1}{2}\be_k\,\frac{1}{2}\int d^3x\, \epsilon _{kij} \Sigma _{ij,0}.
\label{@}\end{equation}
For a particle at rest, the factor to the right of $\be_k$
is the spin vector
$S_k$ of a particle,
so that the interaction energy becomes
\begin{equation}
H_{\rm int}=-\frac{1}{2}{\bf \be}\cdot{\bf S}.
\label{@}\end{equation}
This looks just like the interaction energy
of a unit magnetic moment
with a constant magnetic field, and for that reason we shall refer
to a torsion of the type (\ref{@SB})
as {\em magneto-torsion\/}, and the field ${\bf b}$
as {\em torsion-magnetic\/} field. From the Heisenberg equation
$\dot {\bf S}=i[H,{\bf S}]$,
we obtain
the equation of motion for the spin
\begin{equation}
\frac{d}{dt}{ {\bf S}}=-\frac{1}{2}{\bf S}\times {\bf \be},
\label{@spine}\end{equation}
describing a precession
with
frequency $ \omega =|{\bf \be}|/2$.

The microscopic origin of the spin
of the particle is completely irrelevant for this result. The spin,
being the total angular momentum in the particle's rest frame,
is composed of the orbital angular momenta of all
constituents and their spins.
The details of this composition
depend on the actual quantum field theoretic
description of the particle.
A $ \rho $-meson, for instance, has unit spin.
From the hadronic strong-interaction viewpoint of
bootstrap physics,
the unit spin is explained
by
$ \rho $ being a bound state
of a pair of spinless pions
with unit orbital angular momentum.
In quark physics, on the other hand, $ \rho $
is a bound state of a
quark and an antiquark
with zero orbital angular momentum,
with spins coupled to unity.

Thus, in the quark description, the spin of a $ \rho $-meson in a torsion field
(\ref{@SB}) will precess.
Clearly, a theory of gravity with torsion
can only be consistent with particle physics,
if the same precession frequency
if found for
the hadronic description of
$ \rho $ as a bound state of two spinless
pions.

For present-day
theories of gravity with torsion \cite{hehl},
this postulate presents a serious problem.
In these theories, the energy momentum tensor $T^{\mu \nu }(q)$ of a
spinless point particle
satisfies the local conservation law
\begin{equation}
\bar D_\nu T^{\mu \nu}(q)=
D_\nu^*T^{\mu \nu}(q)+2S_{ \kappa }{}^{\mu }{}_{\lambda}(q) T^{ \kappa  \lambda}(q)
=0,~~~
D_\nu^*\equiv D_\nu+2S_\nu,
\label{covLAW2@}\end{equation}
where
$\bar D_\mu$ is the covariant derivative
involving the
Riemann connection
$\bar \Gamma _{\mu \nu  \lambda }
$, and
$D_\mu$ the
covariant
derivative involving the full affine connection
$ \Gamma _{\mu \nu  \lambda }=
 \bar \Gamma _{\mu \nu  \lambda }
 + K _{\mu \nu  \lambda }$.
It is obvious from the
torsionless left-hand part of Eq.~(\ref{covLAW2@}),
and  was
proved in Ref.~\cite{geodesics},
that such a conservation law
leads directly to {\em geodesic\/} particle trajectories for point-like
spinless particles, governed by the equation of motion
\begin{equation}
 \ddot{q}^\nu  +
\bar	  \Gamma _{\lambda \kappa }{}^\nu \dot{q}^\lambda
	   \dot{q}^\kappa =0,
\label{ge}\end{equation}
where $q^\mu(\tau )$ is the orbit parametrized in terms of
the proper time $\tau $.

This motion is not
influenced by torsion.
As a consequence, the spin of a $ \rho $ meson
at rest would {\em not\/} precess
in the two-pion description, in contradiction
with
the
quark-antiquark description.
Since both descriptions are equally
true, we conclude that
geodesics cannot be the correct trajectories of spinless particles.

~\\
{\bf 3.~}The discrepancy can be avoided by
another option for
the trajectories of
spinless particles in this geometry.
These are
the
{\em autoparallels\/}, which obey
an equation of motion  like (\ref{ge}), but with the full affine connection:
\begin{equation}
 \ddot{q}^\nu  +
	  \Gamma _{\lambda \kappa }{}^\nu \dot{q}^\lambda
	   \dot{q}^\kappa =0.
\label{autop}\end{equation}
The conservation law for the energy momentum tensor of a spinless point particle
leading to autoparallel
motion is \cite{kl}
\begin{equation}
D_\nu^*T^{\mu \nu}(q)=0.
\label{covLAW2x@}\end{equation}
In a flat space with torsion,
Eq.~(\ref{autop}) becomes
\begin{equation}
 \ddot{q}^\nu  +
	 2S^{ \nu}{}_{ \lambda \kappa } \dot{q}^\lambda
	   \dot{q}^\kappa =0.
\label{autop2}\end{equation}
Specializing further to a constant magneto-torsion
(\ref{@SB}),
we obtain $\dot q^0$ = const,
and
find
for the spatial motion in euclidean coordinates
the equation
\begin{equation}
\frac{d^2}{dt^2}{\bf x }=- \dot{\bf x}\times {{\bf \be}}.
\label{autop3}\end{equation}
Thus the constant torsion (\ref{@SB}) acts
on the orbital motion of the spinless point particle
just like a Lorentz force.
It is well known from electrodynamics, that this Lorentz force
causes a precession of the orbital angular momentum
of an electron. Its frequency
is determined by the
magnetic moment of the
 {\em orbital\/} motion,
 whose size for a certain orbital angular momentum ${\bf L}$
is
 half as big as
that of a spin ${\bf S}$ of equal size.
The precession frequency following from
(\ref{autop3}) is therefore $ \omega =|{\bf \be}|/2$.
To show this we simply observe that
(\ref{autop3}) follows from a Lagrangian
$L=\dot{\bf x}^2/2+{\bf a}\cdot\dot{\bf x}$,
describing a particle of unit mass moving in a {\em torsion-magnetic\/}
vector potential
${\bf a}={\bf \be}\times {\bf x}$.
The associated
Hamiltonian depending on ${\bf x}$
 and the momentum ${\bf p}=\dot {\bf x}$
reads
\begin{equation}
H=
\frac{1}{2}({\bf p}-{\bf A})^2=
\frac{1}{2}{\bf p}^2 - \frac{1}{2}{\bf \be}\cdot({\bf x}\times{\bf p})
+\frac{1}{8}({\bf \be}\times{\bf x})^2.
\label{@ham}\end{equation}
The smallness of the gravitational coupling
makes torsion small enough
to
ignore the last term.
From the second term written as $-\frac{1}{2}{\bf b}\cdot {\bf L}$
calculate from the Heisenberg equation
$\dot {\bf L}=i[H,{\bf L}]$ the equation of motion for
the orbital angular momentum:
\begin{equation}
\frac{d}{dt}{ {\bf L}}=-\frac{1}{2}{\bf L}\times {\bf \be},
\label{@}\end{equation}
which is the same as Eq.~(\ref{@spine})
for
the spin, leading to the same precession frequency
 $ \omega =|{\bf \be}|/2$.

A similar study can of course by performed
for an {\em electro-torsion\/} field $S_{i0}{}^0=e^i/2$ with
$e^i=\partial _ia^0$,
so that the autoparallel differential equation
(\ref{autop2}) can be rewritten as
\begin{equation}
\frac{d^2}{dt^2}{\bf x }=-{\bf e}- \dot{\bf x}\times {{\bf \be}},
\label{autop3p}\end{equation}
thus extending
(\ref{autop3}) to an analog of the full  Lorentz equation.
The Hamiltinian (\ref{@ham}) contains then an extra
electro-torsion term $a^0$.
This Hamiltonian may be quantized straightforwardly
to obtain a quantum mechanics in the
presence of
electromagneto-torsion fields.
The eikonal approximation of the Schr\"odinger
wave function will describe autoparallel trajectories.

Although the discussion up to this point assumed
constant
electromagneto-torsion fields ${\bf e}$ and ${\bf b}$,
it is easy to convince ourselves
that the final theory
is also valid for space-dependent fields.

Let us compare this with the
couplings in proper magnetism,
where
in analogy to the universal coupling
of an electric field to the charge
of a particle,
a magnetic field ${\bf B}$ couples universally to the magnetic moments.
For orbital angular momenta
and spin, however,
the magnetic coupling is nonuniversal.
Consider atomic electrons.
They have a
gyromagnetic ratio
 $g=2$
caused by the Thomas precession, so that
the magnetic interaction Hamiltonian is
(ignoring the anomalous magnetic moment)
\begin{equation}
H_{\rm int}=-\mu_B\,{\bf B}\cdot \left({\bf L}+2{\bf S}\right),~~~~\mu_B\equiv \frac{e}{2Mc}
\label{@Zeeman}\end{equation}
where $\mu_B$ is the Bohr magnetic moment (using $\hbar =1$).
In a weak magnetic fields, an atom has
an interaction energy
$-g\mu_BB M$, with the
gyromagnetic ratio
$g=1+\left[J(J+1)+S(S+1)-L(L+1)\right]/2J(J+1)$,
where $J$ is the quantum number of the toral spin vector ${\bf J}={\bf L}
+{\bf S}$. This ratio $g$
causes the characteristic level splitting of the Zeeman effect.
~\\~\\
{\bf 4.~}It is useful to set up a new action for a Dirac field
which is
compatible with the proposed universality principle.
In a first step,
consider a
Riemann-flat spacetime with Minkowski coordiantes $x^ \alpha =(x^0,{\bf x})$
and an action
\begin{equation}
{\cal A}=\int d^4x\,\bar \psi(x)\left[
\gamma ^\alpha\left(if_\alpha{}^ \beta \partial _\beta-eA_\alpha
-\frac{1}{2}K_{\alpha \beta  \gamma } \Sigma^{ \beta  \gamma }\right)
 -M\right] \psi(x), ~~~~~
\label{@ac0}\end{equation}
 where
$f_\alpha{}^ \beta =1-
a_\alpha{}^ \beta $, with  $a^{\alpha}{}^ 0 $ being the
electromagneto-torsion
field $(a^0,{\bf a})$, and the
other components $a_\alpha{}^i=0$ vanishing.
 It is a gauge field
whose curl yields the torsion,
$S_{ij }{}^ 0 =(\partial _i a_ j{}^0\,-\,$$
\partial _j a_ i{}^0)/2$.
The action is gauge-invariant under
$a_i{}^0(x)\rightarrow
a_i{}^0(x)+\partial _i \Lambda^0({\bf x}) $ with a simultaneous transformation
$\psi(x)\rightarrow e^{-i \Lambda  ^0({\sbf x})\partial _0}\psi(x)$.
The $4\times 4$-matrices
$ \Sigma _{ \beta  \gamma }\equiv \frac{i}{4}
[ \gamma _ \beta , \gamma _ \gamma ]_-$
are the generators of Lorentz
 transformations,
so that
the spin current density
in
(\ref{@inta}) is
$\Sigma _{ \beta\gamma ,\alpha}
=-\frac{i}{2}\bar \psi[ \gamma _\alpha, \Sigma _{ \beta  \gamma }]_+\psi
$. Here
 $[~.~,~.~]_\mp$ denotes commutator and anticommutator, respectively,
and all quantities have
standard Dirac notation.
Now we use
the Gordon formula
\begin{equation}
 \bar u({\bf p}',s'_3)     \gamma^\alpha u ({\bf p},s_3)
= \bar u({\bf p}',s'_3)
\left[ \frac{1}{2M}(p'{}^\alpha+p^\alpha)+\frac{i}{2M} \sigma^{\alpha \beta}q_ \beta\right]u ({\bf p},s_3)
\label{q-mel}\end{equation}
to calculate
between single-electron states
of small momenta ${\bf p}'$ and ${\bf p}$
with momentum transfer ${q}=p'-p$
the interaction energy
for slow electrons
\begin{eqnarray}
\!\!\!\!\!\!\!\!\!\!\!\!H_{\rm int}&&=\int d^3x\, \bigg[ \frac{e}{M}{\bf A}(x)
 \cdot\left( {\bf p}\!+\!{\bf q}\!-\!{i}{\bf q}\times {\bf  \Sigma }
\right)
\nonumber \\
&&~~~~~~~~~~~~~~+~{\bf a}(x)
 \cdot\left( {\bf p}\!+\!{\bf q}\!-\!{i}{\bf q}\times {\bf  \Sigma }
\right)-Ma^0(x)-\frac{1}{2}{\bf b}\cdot{\bf  \Sigma }
\bigg] e^{-i{\sbf q}{\sbf x}}       ,
\label{@qqq}\end{eqnarray}
where $ \Sigma _i=\frac{1}{2}  \epsilon _{ijk}\Sigma _{jk}$
are the Dirac spin matrices. We have omitted the
external spinors
$ \bar u({\bf p}',s'_3)$ and $ u ({\bf p},s_3)$, for brevity, since
we shall immediately take the limit ${\bf p}'\rightarrow {\bf p}$
where
$ \bar u({\bf 0},s'_3)u ({\bf 0},s_3)= \delta_{s_3's_3} $,~
$ \bar u({\bf 0},s'_3) \Sigma _{ij}u ({\bf 0},s_3)
=  \epsilon _{ijk}(S_k)_{s_3's_3}$,
and $S_k =\sigma _k/2$, with
Pauli spin matrices
$\sigma _k$.
Before going to this limit, we convert ${\bf q} $ into a derivative of
$e^{-i{\sbf q}{\sbf x}}$, then via an integration by parts
into a derivative of ${\bf A}(x)$,
 and using the vector potentials
${\bf A}=\frac{1}{2}\,{\bf B}\times {\bf x}$ and
${\bf a}=\frac{1}{2}\,{\bf b}\times {\bf x}$, we obtain
in the limit ${\bf p}'\rightarrow {\bf p}$ for a slow electron
\begin{equation}
H_{\rm int}\!=\!\int d^3x\, \left[ \alpha_B\,{\bf B}\cdot({\bf L}+2{\bf S})
-\frac{1}{2}{\bf b}\cdot({\bf L}+{\bf S}) -Ma^0
\right]      .
\label{@en2}\end{equation}
Observe that the last spin term in (\ref{@qqq}) has removed precisely half of the
spin term coming from the coupling of torsion to $ \gamma ^i$,
thus leading to the universal coupling
${\bf b}\cdot ({\bf L}+{\bf S})=
{\bf b}\cdot {\bf J}$.
The Hamiltonian (\ref{@en2})
ensures that nonrelativistic electrons
follow the equation of motion
Eq.~(\ref{autop3p}),
thus running
along
autoparallels
(\ref{autop2}).

To complete the analogy with magnetism,
we make the dimension of the magnetotorsion field equal to that of the magnetic field
by defining
\begin{equation}
{\bf b}\equiv \alpha_K\,{\bf B}^K,
\label{@}\end{equation}
with the
{\em torsionmagneton\/}
\begin{equation}
\alpha_K\equiv \sqrt{G}\hbar /2c,
\label{@}\end{equation}
where $G=\hbar c/M_P^2$, and
$M_P$ is the Planck mass
$M_P=2.38962\times 10^{22}\,M$.
The torsionmagneton is the same
factor smaller
than the Bohr magneton.

Note that in present-day gravity with torsion \cite{hehl,GFCM},
the term
$\frac{1}{2}{\bf b}\cdot {\bf L}$ is absent in
(\ref{@en2}),  while
$\frac{1}{2}{\bf b}\cdot {\bf S}$ is present, in violation of our universlity
principle.
~\\~\\
{\bf 5.~}It is obvious how this theory
can be extended to a more general torsion field.
We simply allow for a full $4\times 4$ matrix $f_\mu{}^ \nu (x)$
in the gradient term of the action (\ref{@ac0}).
Then we allow for a nonvanishing Riemann
curvature by introducing a vierbein field
$h_ \alpha {}^ \nu (q)$ which tranforms
the Minkowski space locally to general coordinates
$q^\mu$ with a metric
$g_{\mu \nu }=h_ {\alpha\mu}h^{ \alpha }{}_ \nu $.
This adds to the contortion
the so-called spin connection
$\mathop{K}^h_{ \alpha  \beta  \gamma  }=
\mathop{S}^h_{ \alpha  \beta  \gamma  }
-\mathop{S}^h_{  \beta  \gamma  \alpha }
+\mathop{S}^h_{ \gamma  \alpha  \beta  }$,
where
$\mathop{S}^h_{  \alpha  \beta \gamma }\equiv
h_{  \alpha  }{}^ \mu h_{   \beta }{}^ \nu\left(
\partial _\mu h_\gamma{}_ \nu
-\partial _\nu h_\gamma{}_ \mu \right)/2$,
and arrive at the action
\begin{equation}
{\cal A}=\int d^4x \sqrt{-g} \bar \psi(x)\left\{
\gamma ^ \alpha h_ \alpha {}^\mu\left[if_\mu{}^  \sigma  \partial _ \sigma
-eA_\mu
-\frac{1}{2}
A_{\mu  \beta  \gamma }
\Sigma ^{ \beta  \gamma  }\right]
 -M\right\} \psi(x), ~~~~~
\label{@Dir}\end{equation}
where
$A_{\mu  \beta  \gamma }\equiv {K}_{\mu  \beta  \gamma }
 +{K}^h_{\mu  \beta  \gamma }$, and
indices are freely converted between
Minkowsi and general coordinates via the matrix $ h_\alpha{}^\mu$
and its inverse $h^ \alpha {}_\mu$ (satisfying $h_\alpha{}^\mu h^\alpha{}_\nu= \delta^\mu{} _\nu)$, and between co- and contravarint
via the metric $g_{\mu \nu }$ and its inverse $g^{ \mu \nu }$.
We have also found it convenient
to introduce a third group of indices
$ \sigma ,\tau , \omega ,\dots $, so that
we may define an inverse $f^{\mu}{}_\tau $
by $ f^{\mu}{}_ \sigma
f_{\nu}{}^ \sigma = \delta^\mu{}_ \nu $.
This action has the usual gauge invariance under local coordinate transformations
and local rotations \cite{hehl,GFCM}.
The nontrivial transformations of the
derivative term can be absorbed in the respective
gauge fields $h_ \alpha {}^\mu$ and
$A_{\mu \alpha  \beta }$.
The covarant curl of $A_{\mu \alpha  \beta }$
is the Cartan curvature tensor.

The action
(\ref{@Dir}) satisfies
our universality principle.
In a Riemann-flat space with electromagnete-torsion,
this is ensured by our very construction
As another example, consider
a space with Riemann curvature and torsion at {\em zero\/} Cartan curvature.
Then $A_{\mu \alpha  \beta }\equiv 0$, i.e., ${K}_{\mu  \beta  \gamma }
 =-{K}^h_{\mu  \beta  \gamma }$, or
\begin{equation}
S_{ \alpha  \beta  }{}^  \gamma
=-h^  \gamma {}_ \nu\left(h_\alpha {}^ \mu\partial _ \mu h_  \beta  {}^  \nu-h_ \beta  {}^ \mu\partial _ \mu h_ \alpha  {}^  \nu \right)/2
=h_ \alpha {}^\mu h_ \beta {}^ \nu
\left(
\partial _\mu h^ \gamma {}_ \nu
-\partial _\nu h^ \gamma {}_ \mu\right)/2.
\label{@curl2}\end{equation}
This is a nonlinear version
of the same type of electromagneto-torsion
as considered above.
Since $A_{\mu \alpha  \beta }\equiv 0$, there
is no direct coupling of the Dirac field to spin.
Then, by the universality principle,
there should also be no coupling to spin
from the derivative
term proportional to $ \gamma^i$.
This is indeed true if we define the
nonlinear relation between gauge field of torsion $f_\mu{}^ \nu $
in the same way as in  Eq.~(\ref{@curl2}),
except with the matrix $h$
replaced by $f^{-1}$.
Then the torsion in a Cartan-flat
space has a gauge field $f=h^{-1}$, so that
torsion completely disappears
from the Dirac action (\ref{@Dir}).
 ~\\~\\
{\bf 6.~}
Note that
also in a torsionless spacetime,
the universality of orbital angular momentum and spin
is satisfied. Then $A_{\mu \alpha  \beta }=\mathop{K}^h_{\mu \alpha  \beta }$
and $f_\mu{}^ \sigma = \delta_\mu{}^ \sigma $, and for a nearly flat $h_ i {}^0=\tilde a^i$ the gradient term
in (\ref{@Dir}) gives
a coupling
$\tilde{\bf b}\cdot({\bf L}+2{\bf S})$,
with $\tilde{b}_i
= \epsilon _{ijk}\partial _j h_i{}^0$,
while the spin term removes
half the spin coupling just as in
(\ref{@en2}), thus leading
once more
to the universal
form $\tilde {\bf b}\cdot {\bf J}$.

Indeed, this can be observed
in a well-known result of Einstein's theory
of garvitation
on
the
gravitational field of a spinning star
which exerts a rotational drag upon a distant
point particle (Lense-Thirring effect).
The deviation of the metric from the Minkowski form
is
\begin{eqnarray}
  \phi ^{00}({\bf x}) & = &
 -4 \frac{GM}{c^2r} + \dots~,~~~~~~ \phi ^{ji} ({\bf x})  =  0,
\nonumber \\
	\phi ^{0i} ({\bf x}) & = & \phi ^{i0} ({\bf x})  =
		   2\frac{G}{c^3r^3} ({\bf x} \times {\bf J})
	      + \dots~ ,
	\label{5.86}\end{eqnarray}       %
where $ G $ is the
gravitational constant, $M$ the mass,
and
${\bf J}$ the total angular momentum of the star at the origin,
obtained
from
the spatial integral over the star volume  $V$:
\begin{eqnarray}
 J^k=\frac{1}{2} \epsilon _{ijk} J^{ij} =
\frac{1}{2} \epsilon _{ijk}\int_V d^3x [x^i{T}{}^{j0} ({\bf x},t) -
	      x^j T^{i0} ({\bf x},t)].
\label{5.73}\end{eqnarray}
The energy-momentum tensor on the right-hand side receives
contributions from both orbital as well as spin
angular momentum.
Thus, a nonrotating polarized neutron star
with total spin ${\bf S}$ gives rise to
the same Lense-Thirring effect as a rotating star
composed of spinless dust
with purely orbital angular momentum ${\bf L}$
equal to the spin ${\bf S}$ of the neutron star.

Conversely, a spinning test particle in an external gravitational field
$g_{\mu \nu }(q)$ is only coupled via its total energy-momentum tensor $T^{\mu \nu }(q)$.
In the rest-frame of a particle, the off-diagonal matrix elements
of $T^{\mu \nu }(q)$ receive equal contributions from
orbital and spin angular momenta.
~\\~\\
{\bf 7.~}In conclusion we see that
only autoparallel trajectories comply with the universality
principle of orbital and spin angular momentum.
This principle guarantees
our ability
to predict
the gravitational behavior of particles
without detailed knowledge on the
source of their spin in terms of its fundamental constituents.
In fact, this knowledge will probably never be
available completely, since every new generations of physicists
discovers additional fundamental particles.

The theory
of a Dirac field defined by the action (\ref{@Dir})
leads to a consistent  description of an alectron in a
gravitational field in which the torsion is restricted to the
to be a curl.
It will be useful to study the
conservation law for the energy momentum tensor
of the electron field following from Eq.~(\ref{@Dir})
and show
that the resulting trajectories
are autoparallels.
The eikonal approximation
to the wave propagation should be performed to confirm this.

The present theory is a nontrivial extension
of the theory with torsion of the gradient type \cite{kl,klpel}.
The remaining eight torsion components will need further
work.

Let us end by remarking that
autoparallel equations of motion
can be derived from the standard action of a classical point particle action
via a modified variational procedure \cite{fiz,pel,PI} which follows from
geometric considerations (closure failure of parallelograms
in the presence of torsion).
The geometric basis for these developments
was derived from an analogy
of these spaces with
a crystal with defects, which
 in crystal
play the same geometric role as
curvature and torsion in gravity \cite{GFCM}.
A {\em nonholonomic mapping principle\/} was
found \cite{kl,PI}
to transform equations of
motion from flat space
to spaces with curvature and torsion.
This was a necessary step in solving another fundamental problem,
the path integral of the hydrogen atom \cite{PI}.

Autoparallel trajectoris are also the most natural trajectories
obtained from an
 embedding of spaces with torsion in a
Riemannian space \cite{embed}.

~~\\Acknowledgement:\\
The author thanks Dr. A. Pelster for useful discussions.

\end{document}